# Interactive Graph Visualization and Teaming Recommendation in an Interdisciplinary Project's Talent Knowledge Graph

**Xu, Jiawei** School of Information, University of Texas at Austin | jiaweixu@utexas.edu
**Chen, Juichien** School of Information, University of Texas at Austin | paulapc058@gmail.com
**Ye, Yilin** Columbia University | yy3152@columbia.edu
**Sembay, Zhandos** School of Medicine, University of Alabama at Birmingham | zsembay8@uab.edu
**Thaker, Swathi** School of Medicine, University of Alabama at Birmingham | snthaker@uab.edu
**Payne-Foster, Pamela** School of Medicine, University of Alabama at Birmingham | pfoster@ua.edu
**Chen, Jake** School of Medicine, University of Alabama at Birmingham | jakechen@uab.edu
**Ding, Ying** School of Information, University of Texas at Austin | ying.ding@ischool.utexas.edu

## ABSTRACT
Interactive visualization of large scholarly knowledge graphs combined with LLM reasoning shows promise but remains under-explored. We address this gap by developing an interactive visualization system for the Cell Map for AI Talent Knowledge Graph (28,000 experts and 1,179 biomedical datasets). Our approach integrates WebGL visualization with LLM agents to overcome limitations of traditional tools such as Gephi, particularly for large-scale interactive node handling. Key functionalities include responsive exploration, filtering, and AI-driven recommendations with justifications. This integration can potentially enable users to effectively identify potential collaborators and relevant dataset users within biomedical and AI research communities. The system contributes a novel framework that enhances knowledge graph exploration through intuitive visualization and transparent, LLM-guided recommendations. This adaptable solution extends beyond the CM4AI community to other large knowledge graphs, improving information representation and decision-making. Demo: https://cm4aikg.vercel.app/

## INTRODUCTION
An academic graph, with nodes representing researchers, papers, and other entities, and edges depicting authorship, collaborations, or usage information, can capture the complex relationships within and across scientific communities (Birkle et al., 2020; Burnham, 2006; Kinney et al., 2023; Wang et al., 2020). Graph visualization tools such as Sci2 Tool (Börner, 2011), VOSviewer (Van Eck & Waltman, 2010), and Gephi (Bastian et al., 2009), and Cytoscape (G. Su et al., 2014) help users effectively explore these complex relationships. For example, users can understand their relative position within the scientific community and identify potential teaming opportunities. However, these tools fall short in interactively displaying large-scale knowledge graphs and offering cross-platform capabilities.

This paper presents an interactive graph visualization solution using an interdisciplinary research project: the Bridge2AI - Cell Maps for AI Data Generation Project Knowledge Graph (CM4AI KG) (H. Xu et al., 2023; J. Xu et al., 2025) as an example. The Bridge2AI-CM4AI project (Clark et al., 2024; Suran, 2022) assembles professionals and researchers from biomedicine, computer science, and social sciences to develop and curate high-quality biomedical datasets critical for AI-driven research and transformative healthcare advancements. The CM4AI KG visualization is designed to identify suitable potential users for these datasets and locate relevant collaborators within the extensive biomedical domain. The tool supports the exploration of the distribution of CM4AI researchers and datasets within the broader biomedical and genomics landscape, aiding the discovery of future collaborators and users. Furthermore, this visualization utilizes language model agents (Y. Su et al., 2024) to offer interactive suggestions for potential collaborators and users for each researcher and biomedical dataset in the CM4AI KG.

Our visualization application is highly adaptable, incorporating node representations via SPECTER2, a scientific document embedding model (Singh et al., 2022), dimensionality reduction techniques such as t-SNE and UMAP (McInnes et al., 2020; van der Maaten & Hinton, 2008), and large-scale responsive node visualization through PixiJS (Van der Spuy, 2015). Compared to widely used graph visualization tools, such as Gephi (Bastian et al., 2009) and Cytoscape (G. Su et al., 2014, p. 3), PixiJS efficiently handles and visualizes large graph datasets while offering excellent cross-platform capabilities. Additionally, we incorporate LLM agents to enhance user understanding and provide explainability. This interactive visualization framework can be adapted for other medical domain knowledge graphs, offering an interactive interface that improves user accessibility. The interactive graph visualization of the CM4AI KG provides an intuitive, user-friendly interface for exploring the knowledge graph.

## DATA AND METHODS

### Data Preparation

The CM4AI KG compiles data from 121 core researchers participating in the Bridge2AI project, including 35 core members of CM4AI. The CM4AI KG uses ORCID identifiers (Haak et al., 2012) for these researchers to ensure accurate author disambiguation. The data was extracted from the PubMed Knowledge Graph (J. Xu et al., 2024) and the Semantic Scholar Dataset (Kinney et al., 2023). It includes 2 million papers (as of 2023), 44,000 authors, and 1,179 biomedical datasets. We filtered out authors with fewer than two publications and no publications since 2020, resulting in approximately 28,000 author nodes for relevance and ease of visualization.

### Author and Dataset Representation

To represent authors and datasets, we used Specter2, a state-of-the-art scientific document embedding model (Singh et al., 2022), to transform the titles and abstracts of the 2.05 million papers into 768-dimensional embedding vectors. For author expertise representation, we aggregated paper embeddings into an author expertise embedding, weighing the author's position in each publication. The first and last authors received a weight of 1, while the k-th author was assigned a weight of 1/k. Authors beyond the 10th position were assigned a uniform weight of 1/10. Dataset representations were obtained by aggregating embeddings from papers that utilized the respective datasets. To identify potential future collaborators for each author, we selected the top 30 similar but previously unconnected researchers by computing the cosine similarity of the embeddings. Similarly, for each dataset, we identified the top 150 researchers who had not used the dataset but had the highest similarity as potential users. In addition to similarity-based recommendations, we developed a multi-LLM agent-powered chatbot that allows users to input their teaming needs and receive potential future collaborators. As an author may have multiple areas of expertise, we cluster the author's publications to obtain multiple expertise embeddings for fine-grained recommendations.

We choose the GPT-4o series (OpenAI et al., 2024) as the backbone models to provide justifications for these recommendations. For collaborator recommendations, we input the model with five recent and five most-cited papers since 2017 (if possible) for each author, along with metadata such as title, journal, citation count, and publication year. The model then generated justifications highlighting the potential benefits of the specific collaboration. For dataset user recommendations, we provided the model with an author's recent and highly cited papers and accompanying dataset descriptions, prompting it to justify why these users should consider the dataset.

### Visualization

For data visualization, we utilized PixiJS (Van der Spuy, 2015), a versatile 2D WebGL renderer capable of efficiently displaying large numbers of nodes in a two-dimensional space. Unlike traditional graph visualization tools such as Gephi (Bastian et al., 2009), Cytoscape (G. Su et al., 2014), Sci2 Tool (Börner, 2011), and VOSviewer (Van Eck & Waltman, 2010), PixiJS leverages WebGL technology to handle and render a large number of nodes, ensuring high cross-platform compatibility. This allows users to explore visualizations effortlessly through any modern web browser. The web application was developed using TypeScript and Svelte, with visualization and hosting code adapted from the open-source anime recommendation project, Sprout (Primozic, 2022). To ensure a well-organized layout, crucial for effective embedding visualization and user experience, we experimented with various dimensionality reduction techniques and layout configurations using Emblaze (CMU Data Interaction Group, 2021). This tool facilitates the visual comparison of embedding spaces and offers built-in dimensionality reduction methods, such as t-SNE and UMAP (McInnes et al., 2020; van der Maaten & Hinton, 2008). These methods were used to condense the 768-dimensional embeddings of authors and datasets into 2D coordinates.

### Expertise-gap Based Teaming Recommendation

To help users explore the knowledge graph and find meaningful connections within the community, we designed an expertise-gap based teaming recommendation system (Figure 2). This system goes beyond similarity-based future collaborator recommendations. Users can either view recommendations based on a recent publication or input their teaming needs directly into the chatbot. The teaming system comprises multiple LLM agents: an expertise gap detection agent, which generates queries for retrieval in the vector database based on the user's previous publications and input, and a reranking agent, which scores and provides justifications for the candidates returned after the initial search. Users can interact with the agents in the left chatbox, and the recommended authors will update in real-time on the right, along with justifications for each candidate collaborator. The system also displays the shortest path for the current user to connect with the candidate author through the co-authorship network.

## USE CASES

The CM4AI KG semantic space visualizes talents, datasets, and bio entities as nodes within a two-dimensional space, as illustrated in Figure 1. Node size reflects the number of publications linked to each talent (log scale), while node shapes and colors distinguish between node types: dataset and bio entity nodes are depicted as squares, and talent (author) nodes as circles. This interactive interface allows users to drag, scroll, zoom, hover over, filter, and search for nodes across approximately 28,000 talent nodes, 1,179 dataset nodes, and 5,000 bio entity nodes. The system offers two main functionalities: (1) Exploration of each user's existing positions in the knowledge space. (2) Recommendation of potential collaborators or dataset users with justifications, including similarity-based recommendations and an interactive system that enables expertise-gap based teaming recommendations. Users can directly input their teaming needs to find collaborators. The demo is available at https://cm4aikg.vercel.app/

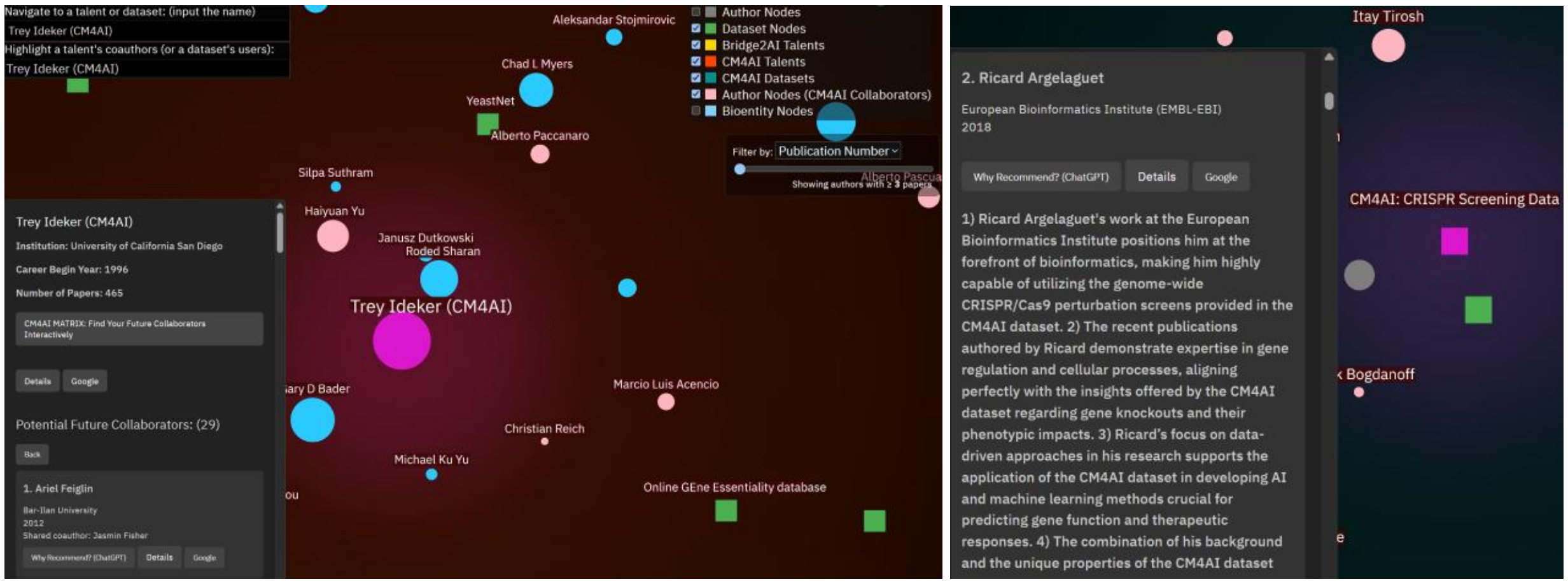

**Figure 1: Overview of the graph visualization. Left: Information window for the talent, Trey Ideker. Users can freely explore the talent space. Right: Recommendation justification for potential future user "Ricard Argelaguet" for the dataset "CM4AI: CRISPR Screening Data" after clicking on "Why Recommend?"**

### Explore Existing User Position in the Knowledge Space

Users can efficiently locate and explore a specific talent (including themselves) or dataset by entering their name in the search box at the top left (Figure 1). The system then provides a list of candidates. Once selected, such as 'Trey Ideker' shown in Figure 1 (left), the view zooms to this talent, presenting an information window with key details such as institution, number of publications, and career start year. Clicking the 'Detail' button directs users to an Oracle APEX interface to review detailed publication histories. This process works similarly when searching for a dataset name. The 'Explore Existing Collaborators or Users' feature helps users understand current collaboration and usage patterns. By entering a talent's name, such as 'Trey Ideker,' in the second search box, users can highlight the nodes of talents who have previously collaborated with that individual in blue, creating a visually distinct effect. This facilitates comparisons between historical and potential future collaborations. The system also provides a node type selection function and filtering options based on the number of publications and career start year.

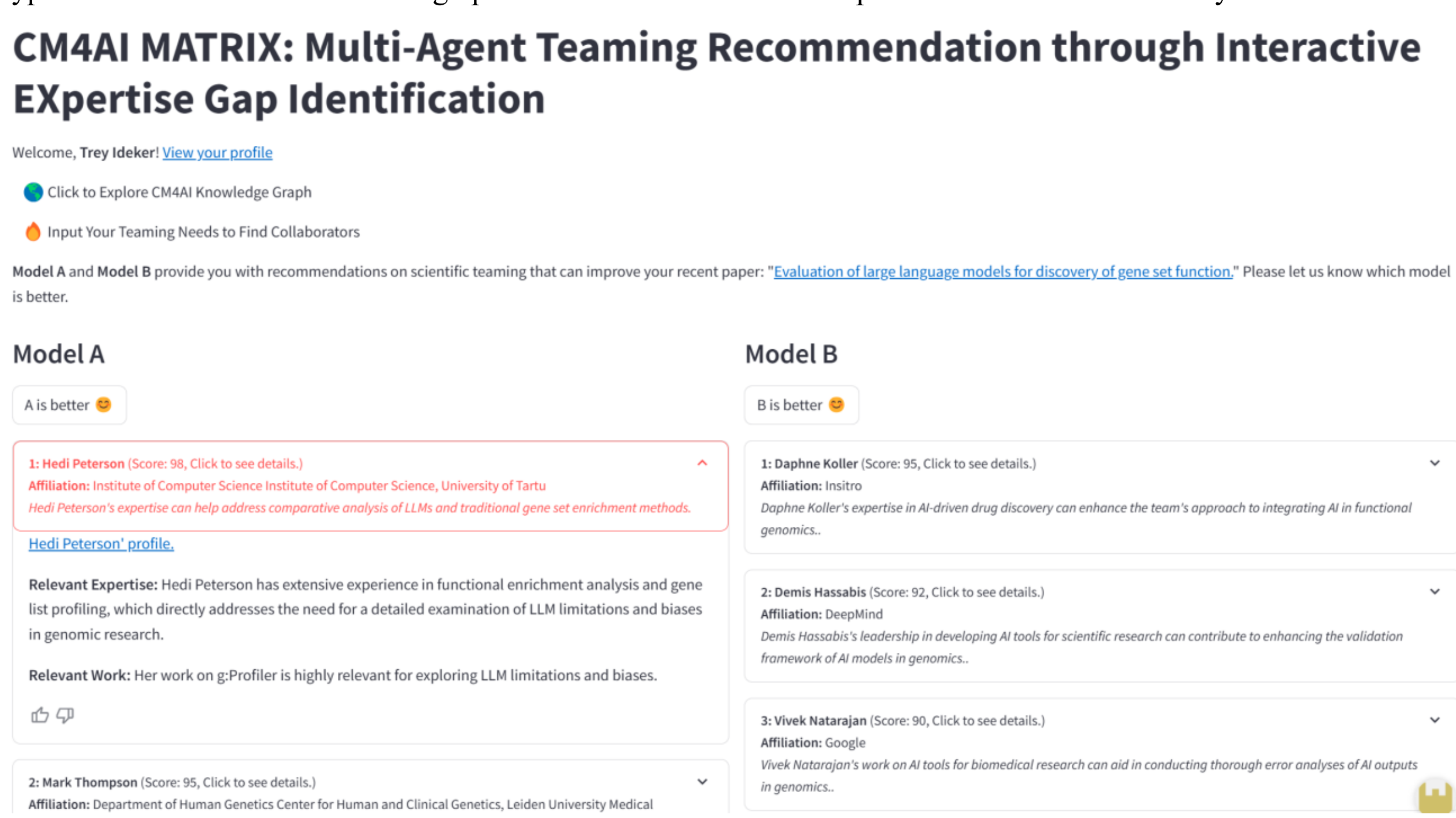

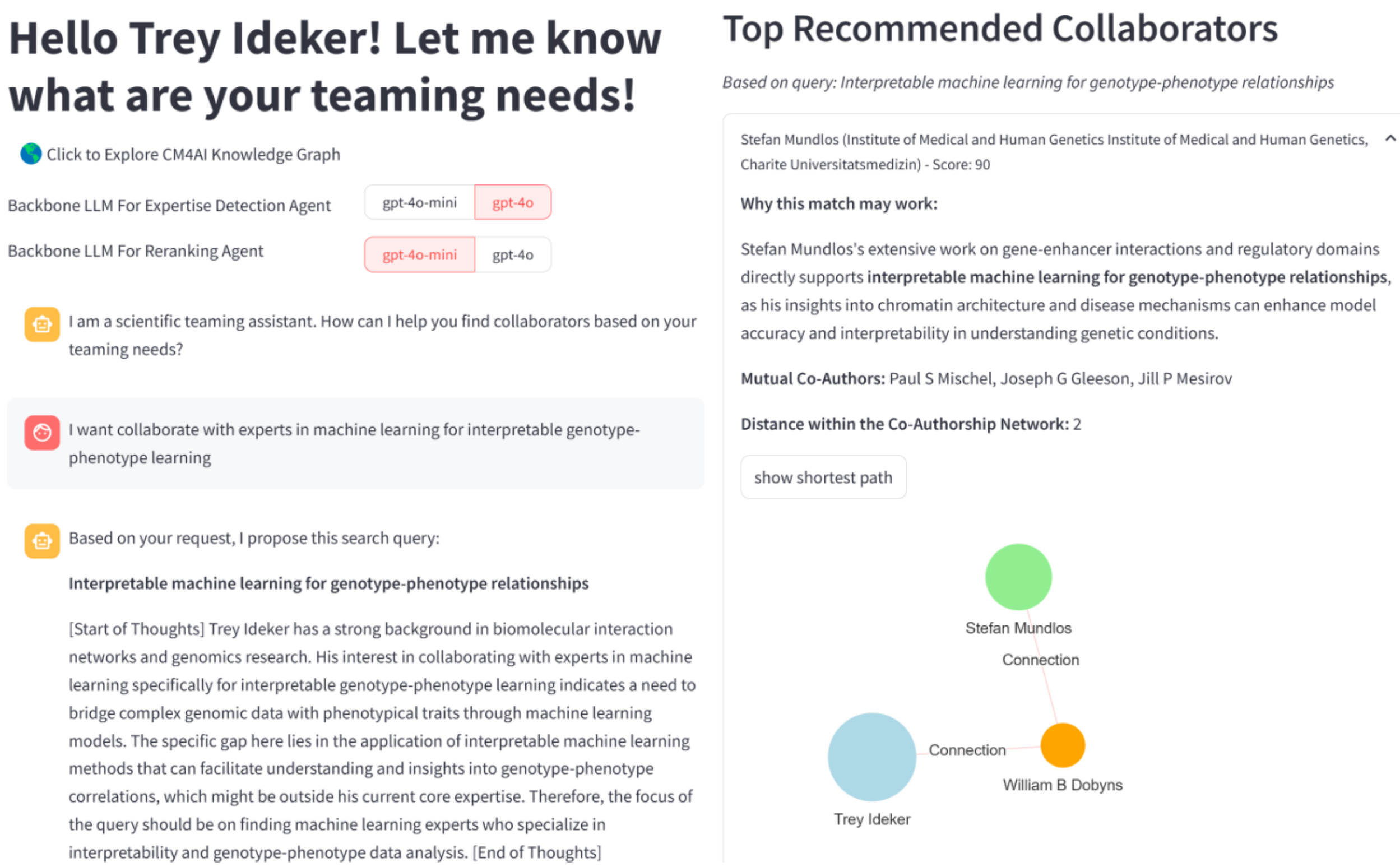

**Figure 2: Expertise-gap based teaming recommendation. After clicking "Find your future collaborators interactively", two customized collaborator recommendation options are available. Top: Recommendation based on the user's recent publication. Bottom: Recommendation based on the user's input.**

## Expertise-Gap Based Interactive Teaming Recommendations

The system provides lists of similarity-based recommended collaborators or dataset users, accompanied by LLM-generated justifications explaining why these individuals are suggested. In Figure 1, users can view these justifications by clicking the 'Why Recommend?' button. For instance, Figure 1 (right) shows the model justifying the recommendation of researcher Ricard Argelaguet for the dataset "CM4AI: CRISPR Screening Data." Beyond similarity-based recommendations, the visualization system offers expertise-gap based interactive recommendations, as shown in Figure 2. When users click "Find Your Future Collaborators Interactively" in the information window (Figure 1), the system navigates to the expertise-gap based interactive recommendations page (Figure 2). Specifically, Figure 2 (Top) displays recommendation results based on the user's specific paper. For example, recommendations for Trey Ideker aim to improve the paper "Evaluation of large language models for discovery of gene set function." These results are generated through the LLM-based multi-agent workflow described previously. Users can also directly input their needs into the teaming chatbot (Figure 2, Bottom), where the expertise gap detection agent and reranking agent collaborate to identify user needs and rank the relevance of each candidate. The system considers whether the researcher can address the expertise gap and displays candidates with justifications, including the shortest path within the co-authorship network to enhance the feasibility and transparency of the teaming recommendations.

## CONCLUSION

The graph visualization provides a rapid response interface for users to interact with the CM4AI Talent Knowledge Graph (CM4AI KG) and offers teaming recommendations. It features high configurability, large-scale knowledge graph visualization (over 30,000 nodes), rich interactivity (search, drag, filter, etc.), and an LLM-powered teaming recommendation system. This showcases the visualization power of WebGL technology and the reasoning abilities of language model agents. Overall, this visualization solution is generalizable beyond the specific CM4AI knowledge graph. It offers considerable advantages for applications that require visualizing extensive datasets and customizing frameworks to incorporate LLM reasoning. By continually refining these capabilities, we can enhance the utility of large-scale knowledge graphs or vector (embedding) data. This advancement can drive progress in medical AI and research by providing user-friendly information representation, thereby facilitating better data interpretation and decision-making.

## ACKNOWLEDGEMENTS

We acknowledge funding support from the NIH Data Generation Projects under the NIH Bridge to Artificial Intelligence (Bridge2AI) Program (Award No. NIH OTA-21-008), as well as from the NSF-CSIRO RESILIENCE project Graph Representation Learning for Fair Teaming in Crisis Response (Award No. NSF 2303038).

## GENERATIVE AI USE

We employed ChatGPT, Claude, and the OpenAI API for the following purposes: ChatGPT and Claude served as coding assistants during the development of the visualization application. The visualization application itself utilizes the OpenAI API. The authors assume all responsibility for the content of this submission.

## REFERENCES

Bastian, M., Heymann, S., & Jacomy, M. (2009). Gephi: An open source software for exploring and manipulating networks. *Proceedings of the International AAAI Conference on Web and Social Media, 3*(1), 361–362.

Birkle, C., Pendlebury, D. A., Schnell, J., & Adams, J. (2020). Web of Science as a data source for research on scientific and scholarly activity. *Quantitative Science Studies, 1*(1), 363–376.

Börner, K. (2011). Science of science studies: Sci2 tool. *Communications of the ACM, 54*(3), 60–69.

Burnham, J. F. (2006). Scopus database: A review. *Biomedical Digital Libraries, 3*, 1–8.

Clark, T., Mohan, J., Schaffer, L., Obernier, K., Manir, S. A., Churas, C. P., Dailamy, A., Doctor, Y., Forget, A., Hansen, J. N., Hu, M., Lenkiewicz, J., Levinson, M. A., Marquez, C., Nourreddine, S., Niestroy, J., Pratt, D., Qian, G., Thaker, S., … Ideker, T. (2024). Cell Maps for Artificial Intelligence: AI-Ready Maps of Human Cell Architecture from Disease-Relevant Cell Lines. *bioRxiv, 2024.05.21.589311.* https://doi.org/10.1101/2024.05.21.589311

CMU Data Interaction Group. (2021). Emblaze [Svelte]. *CMU Data Interaction Group.* https://github.com/cmudig/emblaze

Haak, L. L., Fenner, M., Paglione, L., Pentz, E., & Ratner, H. (2012). ORCID: a system to uniquely identify researchers. *Learned Publishing, 25*(4), 259–264.

Kinney, R., Anastasiades, C., Authur, R., Beltagy, I., Bragg, J., Buraczynski, A., Cachola, I., Candra, S., Chandrasekhar, Y., Cohan, A., & others. (2023). The Semantic Scholar Open Data Platform. *arXiv Preprint arXiv:2301.10140.*

McInnes, L., Healy, J., & Melville, J. (2020). UMAP: Uniform Manifold Approximation and Projection for Dimension Reduction (arXiv:1802.03426). *arXiv. https://doi.org/10.48550/arXiv.1802.03426*

OpenAI, Achiam, J., Adler, S., Agarwal, S., Ahmad, L., Akkaya, I., Aleman, F. L., Almeida, D., Altenschmidt, J., Altman, S., Anadkat, S., Avila, R., Babuschkin, I., Balaji, S., Balcom, V., Baltescu, P., Bao, H., Bavarian, M., Belgum, J., … Zoph, B. (2024). *GPT-4 Technical Report (arXiv:2303.08774). arXiv.* https://doi.org/10.48550/arXiv.2303.08774

Primozic, C. (2022). Sprout: An anime recommendation website. [Jupyter Notebook]. https://github.com/Ameobea/sprout

Singh, A., D'Arcy, M., Cohan, A., Downey, D., & Feldman, S. (2022). Scirepeval: A multi-format benchmark for scientific document representations. *arXiv Preprint arXiv:2211.13308.*

Su, G., Morris, J. H., Demchak, B., & Bader, G. D. (2014). Biological network exploration with Cytoscape 3. *Current Protocols in Bioinformatics, 47*(1), 8–13.

Su, Y., Yang, D., Yao, S., & Yu, T. (2024). Language Agents: Foundations, Prospects, and Risks. *Proceedings of the 2024 Conference on Empirical Methods in Natural Language Processing: Tutorial Abstracts,* 17–24.

Suran, M. (2022). New NIH program for artificial intelligence in research. *JAMA, 328*(16), 1580–1580.

Torvik, V. I., & Smalheiser, N. R. (2009). Author name disambiguation in MEDLINE. *ACM Transactions on Knowledge Discovery from Data (TKDD), 3*(3), 1–29.

van der Maaten, L., & Hinton, G. (2008). Visualizing Data using t-SNE. *Journal of Machine Learning Research, 9*, 2579–2605.

Van der Spuy, R. (2015). Learn Pixi. Js. *Apress.*

Van Eck, N., & Waltman, L. (2010). Software survey: VOSviewer, a computer program for bibliometric mapping. *Scientometrics, 84*(2), 523–538.

Wang, K., Shen, Z., Huang, C., Wu, C.-H., Dong, Y., & Kanakia, A. (2020). Microsoft Academic Graph: When experts are not enough. *Quantitative Science Studies, 1*(1), 396–413. https://doi.org/10.1162/qss_a_00021

Xu, H., Gupta, C., Sembay, Z., Thaker, S., Payne-Foster, P., Chen, J., & Ding, Y. (2023). Cross-Team Collaboration and Diversity in the Bridge2AI Project. *Companion Proceedings of the ACM Web Conference 2023,* 790–794. https://doi.org/10.1145/3543873.3587579

Xu, J., Xie, Q., Liu, M., Sembay, Z., Thaker, S., Payne-Foster, P., Chen, J., & Ding, Y. (2025). Decoding Patterns of Data Generation Teams for Clinical and Scientific Success: Insights from the Bridge2AI Talent Knowledge Graph. In *Proceedings of the 24th ACM/IEEE Joint Conference on Digital Libraries.* Association for Computing Machinery. https://doi.org/10.1145/3677389.3702535

Xu, J., Yu, C., Xu, J., Ding, Y., Torvik, V. I., Kang, J., Sung, M., & Song, M. (2024). PubMed knowledge graph 2.0: Connecting papers, patents, and clinical trials in biomedical science. *arXiv Preprint arXiv:2410.07969.*